\documentclass[10pt,journal,compsoc]{IEEEtran}
\pdfoutput=1
\ifCLASSOPTIONcompsoc
  \usepackage{cite}
  
\else
 
  \usepackage{cite}
\fi
\usepackage{graphicx}
\usepackage{threeparttable}
\ifCLASSINFOpdf
\else
\fi
\usepackage{amsmath}
\usepackage{comment}
\usepackage{algorithm}
\usepackage{algpseudocode}
\usepackage{pifont}
\usepackage{pgfplots}
\usepgfplotslibrary{units}
\usetikzlibrary{patterns}
\usepackage{siunitx}
\usepackage{color,soul}
\usepackage{relsize}

\pgfplotsset{grid style={dashed,gray}}
\usepackage{url}
\begin{document}

\title{Android Malware Detection using Markov Chain Model of Application Behaviors in Requesting System Services}
\author{
\IEEEauthorblockN{Majid Salehi, Morteza Amini$^*$\thanks{* Corresponding author, Tel:+98-21-66166656} \\ $~$ \newline \\ \begin{small}Department of Computer Engineering, Sharif University of Technology, Tehran, Iran \\ \{masalehi@ce., amini@\}sharif.edu\end{small}} }

%
%

\markboth{Submitted to the IEEE Transactions on Information Forensics and Security}%
{Shell \MakeLowercase{\textit{et al.}}: Bare Demo of IEEEtran.cls for Computer Society Journals}

\IEEEtitleabstractindextext{%
\begin{abstract}
Widespread growth in Android malwares stimulates security researchers to propose different methods for analyzing and detecting malicious behaviors in applications. Nevertheless, current solutions are ill-suited to extract the fine-grained behavior of Android applications accurately and efficiently. In this paper, we propose ServiceMonitor, a lightweight host-based detection system that dynamically detects malicious applications directly on mobile devices.
ServiceMonitor reconstructs the fine-grained behavior of applications based on a novel systematic system service use analysis technique. Using proposed system service use perspective enables us to build a statistical Markov chain model to represent what and how system services are used to access system resources. Afterwards, we consider built Markov chain in the form of a feature vector and use it to classify the application behavior into either malicious or benign using Random Forests classification algorithm. ServiceMonitor outperforms current host-based solutions with evaluating it against 4034 malwares and 10024 benign applications and obtaining 96\% of accuracy rate and negligible overhead and performance penalty. 
\end{abstract}

}

\maketitle

\IEEEdisplaynontitleabstractindextext

%
\IEEEpeerreviewmaketitle

\IEEEraisesectionheading{\section{Introduction}\label{sec:Introduction}}
\IEEEPARstart{A}{ndroid} is the most popular smart phone OS currently being used.  Gartner \cite{gartner} reported that 85\% of smart phone sale in the first quarter of 2016 were Android based devices. Android has brought about many new applications, which have resulted in a complete different level of experiences for the end user. However, it has created concern as to the security and privacy of the information (i.e. contacts, geographical locations, photos, and etc.) being used/shared with the new applications. As a result, and due to its popularity, Android OS has been a major target for new malwares. In fact, it has been reported by AV-Test \cite{avtest}, that over 99 percent of new malicious applications targeting mobile devices are aimed at Android devices.

There have been many approaches \cite{bouncer1,crowdroid,drebin,droidapiminer,droidsift,droidmat,droidminer,iccdetector, madam,monet,patronus} proposed to combat the rise of malwares in android devices. These techniques vary in their approaches to the problem; the techniques that operate outside the device, such as the Google bouncer ~\cite{bouncer1}, to the applications that operate on the end-user device ~\cite{crowdroid,drebin,madam} and provide a malware detection service. It is important to note that these different approaches are complement of each other and it is a good example of defense in depth. This could be explained by the fact that end-user detection is bound by  limited resources available on the device, but more importantly, market based techniques are handicapped by the fact that not all applications are downloaded from a single market. In fact, there are multiple alternative markets to the Google play, such as Amazon app store, SlideME, Samsung Galaxy Apps, etc. This highlights the importance of employing the end-user device malware detection techniques.

Techniques proposed for end-user devices are limited by the resources on the devices as well as the user's patience (due to the overhead of the detection mechanism). Generally, these techniques operate by collecting the behaviors (as a set of features) of the application being analyzed and then decide on the nature of the application (i.e. benign or malicious). The analysis could be done either statically or dynamically. 

Static techniques analyze the application's bytecode with near-complete coverage, considering all execution paths, even if a part of the program never executes. Nevertheless, these techniques are vulnerable to transformation attacks \cite{droidchameleon} and could be evaded by obfuscation techniques such as Java reflection and bytecode encryption. Furthermore, malicious behaviors may be implemented in native codes \cite{native}, which are not analyzed, or hidden and triggered in run-time through additional codes loaded dynamically from external sources ~\cite{coadeload}.

Alternatively, dynamic analysis techniques observe the behavior of applications at run time. But as Zhang et al. \cite{vetdroid} reported, most of these techniques which operate on the end-user devices, consider the behavior of applications at the system call level (i.e. the system calls executed and the order of their execution). In fact, given the Android OS architecture,  such techniques obtain an incomplete view of the behavior of the binary being analyzed. This becomes clear, when one considers that in the Android OS, applications are not able to directly access system resources (e.g. SMS, camera, microphone, etc.) through system calls. In fact, since applications are encapsulated in their own memory space in Android, an IPC (Binder) mechanism is provided through which applications can access system services by a specific system call named \textit{ioctl}. Hence, traditional dynamic methods, which are based on analyzing system calls, are less effective in malware detection, as they reconstruct the behaviors of the applications based on a number of intercepted system calls (e.g. \textit{ioctl}), which contain no information about the system services being accessed.

Considering the above noted observations and taking into account the issues and limitations of static analysis techniques, we propose a novel \textit{System Service Use Analysis} technique to capture and analyze the fine-grained behavior (i.e. at the system services level) of applications with the aim of detecting malicious applications. We build a statistical model to represent what and how system services are used to access system resources. Specifically, we model sequences of requested functions from system services as Markov chains, and use them to extract features and perform classification.
To the best of our knowledge, ServiceMonitor is the first method that systematically analyzes and models system service use at multiple levels of semantics regardless of whether it is from Java or native code, and classifies Android applications as malicious or benign directly on mobile devices.  In summary, this paper makes the following contributions:
\begin{itemize}
	\item \textit{System service use analysis.}  
	We propose a systematic system service use analysis technique, which automatically and seamlessly models the state transitions achieved by the functions requested from system services as a Markov chain; aim at representing the application behavior by its pattern of accesses to system resources.
	\item \textit{Effective Android malware detection.} 
	We developed a dynamic Android malware detection framework, named ServiceMonitor, to reconstruct the behavior of applications based on system service use analysis technique with the aim of identifying malicious behaviors as well as malwares. ServiceMonitor is capable to detect Android malwares with high accuracy and few false positives.
	\item \textit{Lightweight detection procedure.}
	With taking into account the limitation of resources in mobile devices, ServiceMonitor can efficiently apply system service use analysis directly on mobile devices and detect malicious applications in reasonable time.
\end{itemize}	

In the remainder of this paper, we first survey a number of related work in Section \ref{sec:RelatedWork}. Then the proposed system service use analysis technique and ServiceMonitor's detection method are described in Section \ref{sec:OurApproach}. After that we present the implementation details and evaluation results in Section \ref{sec:Evaluation}. We also discuss some related issues on Android malware detection and the limitations of ServiceMonitor in Section \ref{sec:Discussion} and conclude the paper in Section \ref{sec:Conclusion}.

\section{Related Work}
\label{sec:RelatedWork}
 
There have been many techniques suggested in the literature for analyzing and detecting Android malwares. These techniques could be categorized based on how the analysis/detection agent is deployed as: i) Emulator-based techniques, ii) Cloud-based techniques, which collect information from end-user device and then aggregate and analyze the information on the cloud, and 
iii) Host-based techniques, which are deployed wholly on the mobile device. 

\subsection{Emulator-based Analysis and Detection}

Google bouncer \cite{bouncer1} is one of the most well-known systems, which automatically analyzes applications on Android official market and purges malicious apps. Although there is little public documentation available,  based on  a set of experiments presented in~\cite{bouncer2}, Google bouncer is a dynamic analysis system that runs and monitors applications in a QEMU\cite{qemu} based emulator which can be easily detected and evaded. AppsPlayground \cite{appsPlayground}, Andrubis \cite{andrubis}, and DroidBox \cite{droidbox} are alternate analyzers, which are based on the QEMU emulator and extend a popular taint tracking framework named TaintDroid \cite{taintdroid} to track sensitive information with the goal of malware analysis. However, due to the dynamic taint analysis, such approaches are vulnerable to evasion attacks presented in \cite{babil2013effectiveness,cavallaro2008limits,slowinska2009pointless}.

Other efforts, such as DroidScope \cite{droidscope} and CopperDroid \cite{copperdroid1,copperdroid} are  dynamic analysis platforms that leverage virtual machine introspection \cite{vmi} to extract the behaviors of applications considering OS and Android-specific views. Dash et al.~\cite{droidscribe} proposed DroidScribe as a dynamic multi-classification system built upon CopperDroid. DroidScribe classifies malicious applications in known malware families based on extracted behaviors from CopperDroid sandbox. The reader is referred to~\cite{offline}, for a good comparison of dynamic analysis techniques in controlled environment.

All of the mentioned systems are based on software virtualization and emulation techniques. So, they suffer from several fundamental limitations (i.e. performance penalties, transparency issues, and special software requirements) for deploying on end-user devices. Furthermore, all of them are vulnerable to evasion attacks~\cite{towards,fistful,evading,morpheus} by which the presence of virtual machine could be detected by malicious applications.

\subsection{Cloud-based Analysis and Detection}

Another approach proposed by a number of researchers is to have data collected on the end-user device, but offloading the analysis to the cloud in order to obtain an aggregate view of the applications being executed on different devices, and minimize the cost of analysis on each device. For example, Crowdroid~\cite{crowdroid} is a dynamic analysis based approach that collects system call logs employing \textit{strace} tool, and transmits the logs in addition to the information about each device as well as the installed application list to a dedicated server. The collected information is then processed to obtain a set of identifying features with which the malicious applications are detected.

Recently, Sun et al. \cite{monet} proposed Monet, an Android malware detection system, that is based on constructing dependency graphs to model the dependencies between application components and system services. But the analyzing and detection procedure of Monet requires a backend server for graph mining and similarity computation. 
Alternatively, ParanoidAndroid \cite{paranoid} deploys a virtual clone of the given mobile device on a remote server synchronized with activities on the real mobile device. Hence, researchers could perform detection techniques on the collected traces without disrupting end users. However, the effectiveness of crowdsourcing frameworks relies on the reaction of end users when they are asked to send recorded logs from applications to an external server. 

\subsection{Host-based Analysis and Detection}
As an alternate approach, a number of techniques are proposed which are executable on the end-user device given the performance constraints, without requiring emulators or cloud based analyzers.

Kirin~\cite{kirin} is one of the first methods that statically detects malicious Android applications considering requested permissions that break conservative security specifications.   
A number of proposals look at the APIs called by each application, in order to decide on the intent of the application (i.e. malicious or benign). For example, DroidAPIMiner~\cite{droidapiminer} captures the names and parameters of sensitive APIs as features for classifying Android applications. DroidMiner~\cite{droidminer} considers the relations between multiple sensitive APIs and not only their frequency or names. Along similar lines, Zhang et al.~\cite{droidsift} proposed DroidSift, a semantic-based approach that leverages a weighted contextual API dependency graphs to classify Android applications. Unlike other static solutions, DroidSift and DroidMiner are resistant against minor transformation attacks \cite{droidchameleon}; because they extract features based on application's behavior (i.e. dependency of API calls).

Furthermore, there have been a number of static learning based studies in which the features are defined and selected based on some malicious code patterns and heuristics. DroidMat \cite{droidmat} and Drebin \cite{drebin} are two similar approaches that extract detection features from applications disassembled codes and manifest the files as much as possible. These two approaches hold the occurrence of sensitive API calls in their feature sets beside other information like requested permissions and names of application components (i.e. activities, services, ‌broadcast receivers, and content providers). These approaches also differ from each other in how to
structure features extracted from the applications. ICCDetector \cite{iccdetector} is a static based method that extracts ICC (Inter-Component Communication)-related features that hold interactions within or cross applications' components, and then leverage machine learning techniques to perform classification. Actually, the ICC-related features represent the pattern of communications between applications' components, which are done through messages named Intents (i.e. communications in Java abstraction layer).
As noted briefly in introduction and stated in \cite{droidchameleon}, static analysis and detection techniques are thought to be insufficient to detect malware variants generated by transformation and obfuscation attacks. Also, most of the current static solutions are ill-suited to analyze additional codes loaded dynamically from external sources \cite{coadeload} as well as native codes \cite{native}.

On the other hand, there have been a number of proposed works in which applications are analyzed dynamically to detect malwares at runtime.  MADAM  \cite{madam} \cite{madam2} is a hybrid  system that defines several syscall based features beside some user activity based features related to user idleness and sent SMS messages to spot malicious activity on the mobile device directly.
However, as Zhang et al mentioned in \cite{vetdroid}, all syscall based systems share one fundamental limitation. Due to the missing high-level Android-specific semantics, their analysis is ineffective to demonstrate the fine-grained and accurate behaviors of Android applications. 
To solve this limitation, Sun et al. \cite{patronus} proposed Patronus, which focuses on fine-grained behaviors of applications. Patronus is a host based intrusion prevention system for Android devices. Patronus considers a database of manually crafted malicious policies from known malware samples and calculates the similarity score between transaction footprint extracted from running applications on user's device and malicious transactions (recorded in policy database) aim at preventing malicious intrusions and detecting malwares at run-time. 
But Patronus depends on experts to define security policies and rules to cover malware misbehaviors. Hence, due to the known malicious policies in its database, it could not be effective and suited for detecting unknown and zero-day malware families.

In what follows, and based on the noted limitations of the previous related works, we propose an effective dynamic detection method for malware detection on end-user devices.

\section{Our Approach}
\label{sec:OurApproach}

In the Android security architecture, each application is isolated in a separate sandbox, and direct access from applications to the system resources (i.e. SMS, GPS, address book, etc.)  is prohibited. Instead, access to system resources is handled through the binder driver component, which verifies that the given application has the proper permission to access the specific system resources. On the other hand, applications require access to different system resources to operate properly. Where such operations could be benign as advertised by the application creator, or malicious; although appears to be benign by the client. In both scenarios system resources are accessed, but the type of resources and the order of accesses are different.

 \begin{figure*}
 \center
	\includegraphics[width=.9\textwidth]{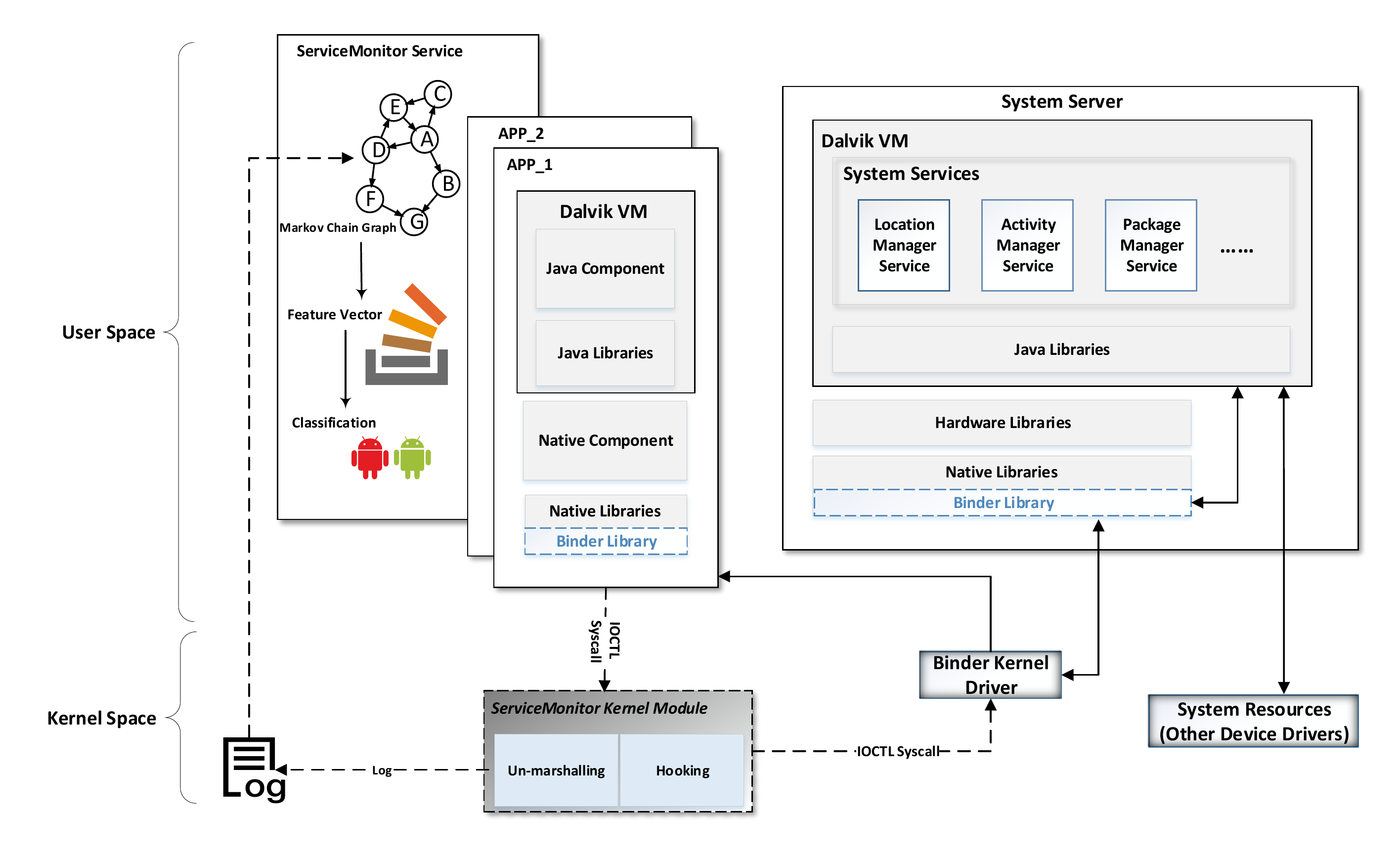}
	\caption{Architecture and detection process of ServiceMonitor.}
	\label{plot:arch}
\end{figure*}
  
The malicious behaviors that are frequently observed from Android malwares have been widely surveyed in the research literature \cite{stowaway,droidapiminer,survey1,survey2,survey3} and malware reports~\cite{contagio}. Based on those studies and the analysis that we have done on both benign and malicious samples, we believe that a fine-grained behavioral model could be constructed through observing accesses to a dozen of system services and their functions implemented in the framework layer of Android operating system.
Furthermore, and based on the Android system documentation~\cite{android}, system services could be categorized into six broad categories. These categories and the services that they cover are shown in Table \ref{tab:features}. 

\begin{table*}[t]
	\renewcommand{\arraystretch}{1.3}
	\caption[SYSTEM SERVICES]{System services could be categorized into 6 categories. The information noted in this table was obtained from documentations available at developer.android.com}
	\label{tab:features}
	\centering
	\begin{tabular}{p{2.5cm}|p{3cm}|p{12cm}}
		\hline
		Service Category  & Interfaces &Functionalities  \\
		\hline
		\hline
		
		Telephony Manager   & ISms \newline IPhoneSubInfo \newline ITelephony \newline ITelephonyRegistry &Sends text messages, retrieves phone numbers, retrieves the unique device ID (e.g., IMEI), retrieves the serial number of the ICC, retrieves the unique subscriber ID (e.g., IMSI), retrieves the software version number of the device (e.g., IMEI/SV), retrieves the network type for data transmission, retrieves the current active phone type (e.g., PHONE\_TYPE\_CDMA, PHONE\_TYPE\_GSM), listens to the phone state changes  \\
		\hline
		Location Manager& ILocationManager  &   Retrieves the last known location, registers for location updates, retrieves the list of the names of LocationProviders that satisfy the given criteria, retrieves the name of the provider that has the most compliance with the given criteria    \\
		\hline
		Network Manager&  IConnectivityManager \newline IWifiManager  &  Retrieves the current proxy settings, retrieves the connection status information of a particular network type, retrieves the details of the current active default data network, retrieves the connection status information of all network types supported by the device,  retrieves the dynamic information about the current Wi-Fi connection, retrieves the Wi-Fi enabled status   \\
		
		\hline
		Activity Manager&  IActivityManager & Starts a service, Stops a service, Resumes an activity, Idles an activity, Gets the list of running application processes, Checks permissions, Retrieves the memory information, Registers for Intent broadcasts (e.g., ‌‌Boot\_Completed), Broadcasts Intents, Gets a content provider, Removes a content provider, Starts an activity, Pauses an activity, Finishes an activity, Gets services, Unregisters Intent receivers, Gets the orientation, Sets the orientation, Kills a list of processes, Gets task id of an activity, Gets the sender of a given Intent  \\
		\hline
		Package Manager&  IPackageManager &  Retrieves the list of all packages installed,  retrieves information about a particular package/application, retrieves the names of all packages that are associated with a particular user id, retrieves information about an application package installed on the system, retrieves information about a particular activity class, retrieves all activities can be performed for the given intent, checks whether a particular package has been granted a particular permission, checks for the presence of the given feature name in OS   \\
		\hline
		OS Related Activities & IPowerManager \newline IServiceManager \newline IMountService&  Retrieves the overall interactive state of the device (i.e.  actual state of the screen),  acquires the wake lock and forces the device to stay on, releases the wake lock, retrieves an existing service with the given name, retrieves the state of a volume via its mount point, retrieves the list of all mountable volumes   \\
		\hline
		
	\end{tabular}
\end{table*}

In what follows, we discuss the details of the detection process in which ``ServiceMonitor" receives the execution traces (i.e. functions requested from system services) of an application and then models the behavior of the application as a Markov chain. Finally, a feature vector is generated from the Markov chain model and fed to a binary classifier to determine the nature of the given application as either malicious or benign. Fig. \ref{plot:arch} represents the mentioned process beside the architecture of the proposed monitoring system.

\subsection{Monitoring of Applications' Service Requests}
As noted earlier, in the Android security model, all transactions to/from each application (including those appears between the application and other applications or between the application and system services) would be only possible through the Binder library. 

The Binder library itself is divided into two segments:
\begin{itemize}
	\item[i)] a user-space shared library called \emph{libbinder.so} and
	\item[ii)] a kernel-level driver.  
\end{itemize}
\emph{libbinder.so} is tasked with receiving requests from the user-space process and marshalling them into a {\em Parcel} object which is then passed to the kernel-level binder driver for further processing, given the permission granted to the application. More specifically, the {\em ioctl} system call is employed by applications to make requests to the Binder library.

In order to obtain detailed information about the system services requested by each application, we implemented a kernel module with which {\em ioctl} system calls are intercepted and unmarshaled (i.e. parsed). More details on the implementation are provided in Section~\ref{sec:Evaluation}. The architecture and overall design of ServiceMonitor is depicted in Fig. \ref{plot:arch}. ServiceMonitor's kernel module is split into hooking and unmarshaling components for obtaining $ioctl$ system calls and unmarshaling them respectively. We should note that the unmarshaling procedure, which is introduced in the following, is based on the previous works \cite{artenstein2014man,copperdroid1,schreiber2011android} that present the structure of $ioctl$ system calls and architecture of Binder transactions. 

The $ioctl$ system call has the following syntax:

\centerline{{\em \textbf{ioctl(Driver\_fd, BINDER\_WRITE\_READ, \&bwr);}}  }
Since we aim at intercepting ‌Binder transactions, {\em /dev/binder} is the only considerable value for us that determines the file descriptor of Binder devices as the first $ioctl$ argument. 

{\em BINDER\_WRITE\_READ} command is the basis for all IPC operations. Thus, we consider that as the request code that should exist in the intercepted {\em ioctl} system calls. 

\begin{figure}
\center
	\includegraphics[width=.9\columnwidth]{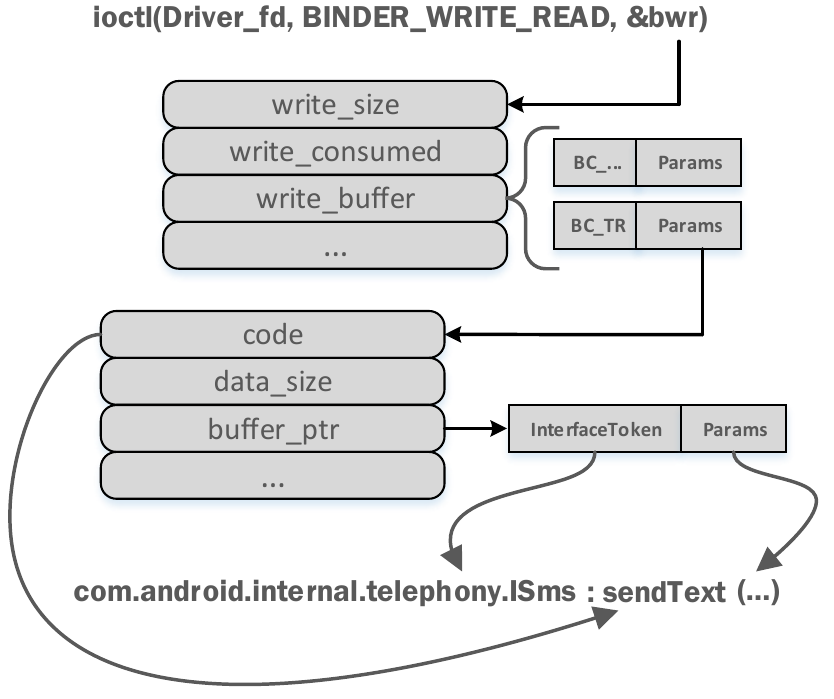}
	\caption{Dissecting \textit{ioctl} syscall. ServiceMonitor extracts the target system service (e.g. com.android.internal.telephony.ISms) and requested function (e.g. sendText) by dissecting \textit{ioctl} system calls.  
	} 
	\label{plot:ioctl}
\end{figure}

The last and the most important argument of {\em ioctl} system call is a pointer to a struct of the type {\em bwr} (short for {\em binder\_write\_read}). 
As illustrated in Fig. \ref{plot:ioctl}, this data structure contains a pointer to a valuable buffer named {\em write\_buffer}, which holds the type of transactions and the respected parameters. Due to the fact that we would like to intercept Binder transactions,  {\em BC\_TRANSACTION} is the only one of these transaction types that is of interest to our work. 
At the next level and in the Binder transactions, we are dealing with a data structure named {\em binder\_transaction\_data}. As depicted in Fig. \ref{plot:ioctl},  this data structure contains some valuable attributes that could be used for extracting system services and corresponding functions that are requested through the invocation of this {\em ioctl} system call. 
The $code$ attribute is the code of the requested function, which is implemented in the destination system service and is required to be executed by the source application.
\textit{buffer-ptr} is a pointer to the $Parcel$ object that we would like to unmarshal. There is a 16-bit Unicode string named \emph{InterfaceToken} at the start of every $Parcel$ object structure.  \emph{InterfaceToken} determines the name of the system service that is considered as the server for the application request.

Finally, the requested functions and system services are recorded in a chronological order as behavioral features and delivered to ServiceMonitor's service for modeling and classification. Owing to the fact that these characteristics of Android OS have stabled during all version releases of Android, our IPC dissecting procedure is a portable way to reconstruct application behaviors, independent from the internal complexity of applications. 

\subsection{Markov-chain Modeling of Application Behaviors}
\label{markov}
ServiceMonitor employs Markov chains for modeling application behaviors. A Markov chain could be described by a weighted directed graph. Actually, in this representation there is a set of nodes that belong to different states, and a set of edges between them weighted with the probability of transition from each node to another one. 

In our proposal, for each application, the potential dependencies between the states obtained by calling functions of system services are represented as a Markov chain. 
In fact, ServiceMonitor generates a complete weighted directed graph in which each vertex of $\varphi$ corresponds to a state obtained by calling a function of a system service (using $ioctl$ system call). For each function $F_x$ of system services, we take an abstract state (denoted by $State(F_x)$) representing the application state after calling the function. In this way, the generated graph has  $|\varphi|^2$ edges and each edge has a weight; representing the probability of the corresponding state transition.

In this modeling scheme, the weights of edges are obtained by analyzing the sequences (traces) of requested functions from system services; which are logged by the monitoring system. The distance between two requested functions $F_i$ and $F_j$ in a sequence  $\sigma$   (where $i$ and $j$ are the indexes of them in the sequence) is defined as follows:
 $$d(F_i , F_j )=j-i$$ 
 Actually, the distance value $d(F_i , F_j )$ can determine the potential relationship (i.e. data or control flow) between a pair of requested functions $F_i$ and $F_j$  in a sequence  $\sigma$. In other words, two requested functions that are closer to each other have more contribution on the weight of the edge between their corresponding states in the Markov chain graph.

Now, we define $P_{xy}$ as the probability of transition from state $s_x$ to $s_y$ (the weight of the directed edge $(s_x , s_y )$) as the following, where $State(F_z)$ determines the abstract state of the application after receiving the requested service function $F_z$.
  \[ FV_{xy} =
 \begin{cases}
 0       & \text{   , if } x=y\\
 & \\
  \mathlarger{\sum}\limits_{\substack{i<j \leq |\sigma|,\\s_x=State(F_i), s_y=State(F_j),\\ \not \exists h, (i<h<j ~\wedge~ s_x=State(F_h)}} \mathlarger{\frac {1}{d(F_i , F_j )}}   & \text{   , otherwise }\\
 \end{cases}
 \]
 
\noindent $P_{xy} =FV_{xy}/ \mathlarger{\sum}\limits_{\substack{1\leq l \leq |\sigma|}} FV_{xl} $
 
Note that in this paper, for the sake of simplicity, we label the state of an application after receiving a requested service, by the name of the requested function from system services (e.g., see Fig. \ref{plot:markov}; which is the Markov chain model of the example represented in the next section).   
 
The pseudo-code of building the Markov chain graph is represented in Procedure \ref{algorithm}.
 
\floatname{algorithm}{Procedure}
\renewcommand{\algorithmicrequire}{\textbf{Input:}}
\renewcommand{\algorithmicensure}{\textbf{Output:}}
\begin{algorithm}
	\caption{Building the Markov chain graph}
	\label{algorithm}
	\begin{algorithmic}[1]
		\Require system\_service\_trace, system\_service\_list
		\Ensure  probability\_matrix
		\State size=len(system\_service\_list)
		\State Declare Integer FV[size][size]
		\State j=len(system\_service\_trace)
		\For{$i=0 \rightarrow j-1$ }
		\State line=system\_service\_trace[i]
		\State index=system\_service\_list.index(line)
		\State append index to map\_list
		\EndFor
		\State k=len(map\_list)
		\For{$i=0 \rightarrow k-1$ }
		\For{$j=i+1 \rightarrow k-1$ }
		\If {$map\_list[i] \neq map\_list[j]$}
		\State FV[map\_list[i]][map\_list[j]]+=(1/(j-i))
		\Else
		\State Break
		\EndIf
		\EndFor
		\EndFor
		\For{$i=0 \rightarrow size-1$ }
		\State S=SUM(FV[i][:])
		\For{$j=0 \rightarrow size-1$ }
		\State FV[i][j]=FV[i][j]/S
		\EndFor
		\EndFor
		\State Return FV
	\end{algorithmic}
\end{algorithm}


\subsection{Feature Extraction and Classification}
The final step for determining the nature of an application is extracting a feature vector and feeding the vector into a machine learning algorithm for classification of the application behavior. To this aim, we take the weights of $|\varphi|^2$ edges of the Markov chain graph (generated from the logged sequences of requested functions from system services) as a 1D feature vector $fv = [f_1, f_2, ..., f_{|\varphi|^2} ]$, where $f_i$ is equal to $P_{km}$ so that $i = (k - 1) \cdot |\varphi| + m$.

Then we employ Random Forests algorithm for classifying the application behavior to either malicious or benign using the extracted feature vector. To train the classification algorithm, we can use the samples of known malwares and benign applications. Section \ref{sec:Evaluation} describes the training, evaluation, and feature reduction process in more details.

\begin{figure}
	\center
	\includegraphics[width=0.37\textwidth]{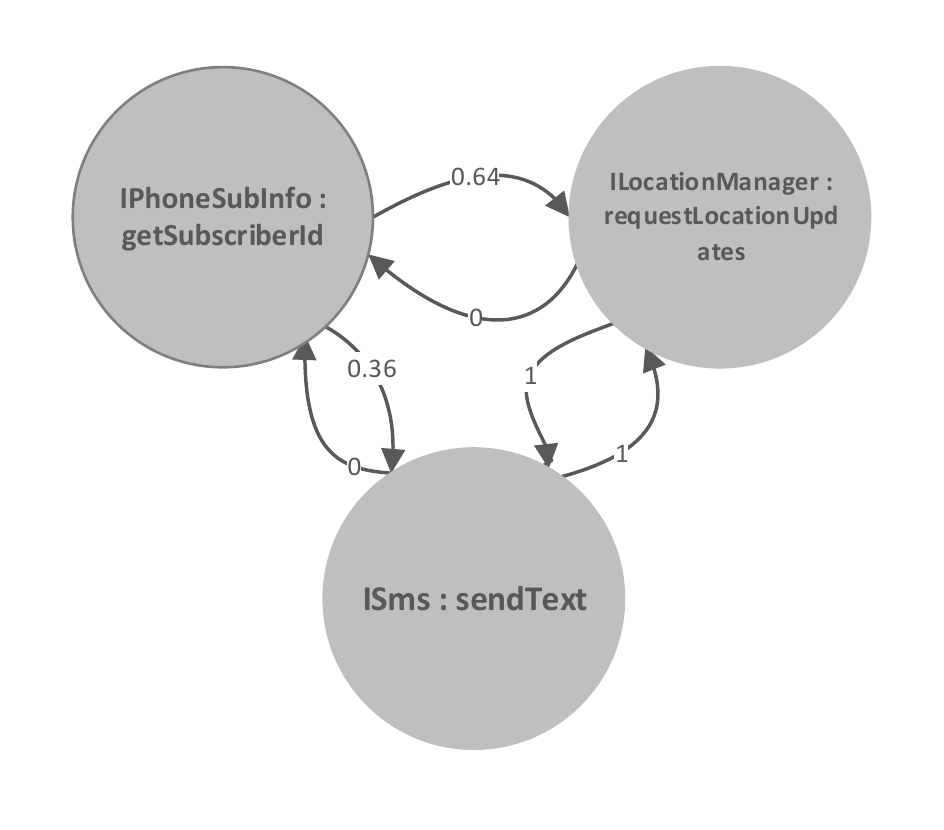}
	\caption{Markov chain model of an application's behavior.}
	\label{plot:markov}
\end{figure}

\textbf{Example:}
Consider an application that requests the following functions respectively: \emph {getSubscribedID, requestLocationUpdates, sendText, requestLocationUpdates, sendText}. Following the proposed approach, we monitor and extract these functions requested from system services by our proposed IPC dissecting procedure and model them in a Markov chain graph like Fig. \ref{plot:markov}. Then, we extract the specified feature vector as [0, 0.64, 0.36, 0, 0, 1, 0, 1, 0] from the Markov chain model of application behavior for classification purpose. Note that in this example we supposed that we have only three functions in Android architecture.

\section{Implementation and  Evaluation}
\label{sec:Evaluation}
We implemented our proposed method in a system called ServiceMonitor. 
The main module is a kernel module, which is implemented in \emph{C} and is tasked to monitor each application and log the functions requested from system services by each application in their chronological order. More specifically, the developed kernel module intercepts the $ioctl$ system calls and dissects them to extract the functions requested from the system services. This is done by rewriting the address of the $ioctl$ function implementation in \textit{system\_call\_table}. 

\textit{system\_call\_table} is implemented in kernel level and is used for organizing system functions and quick access to them. Hence, at first, we need to obtain the address of \textit{system\_call\_table} from the \textit{vector\_swi} handler. Indeed, by using the technique proposed in \cite{hooking}, we are able to obtain the address of \textit{system\_call\_table} in all versions of Android OS in a similar fashion. After that, we rewrite the address of the $ioctl$ function and redirect all Binder transactions to our unmarshaling component. Finally, after logging and dissecting Binder transactions, we redirect them to the original $ioctl$ function for continuing the normal execution of the application. 

The collected features are then processed using a second module, implemented in Python, with which the requested system services are transformed into the Markov chain representation discussed earlier. 

Lastly, an implementation of the classification algorithm, based on the randomForest library \cite{randomforest_imp} written in $R$ language \cite{r}, is used to generate a model for distinguishing the malicious applications from the benign ones. The introduced modules are collected in a standalone application, which could be deployed on an end device with the  aim of employing a detection model, which is trained offline, and classifying running applications as either malicious or benign. The source code of the above described modules is available at \url{http://ce.sharif.edu/~masalehi/ServiceMonitor}. 
 
In what follows, we first describe the dataset used in evaluating the proposed framework in Section~\ref{dataset}. The details on how the features are extracted from the collected dataset are presented in Section~\ref{fcollect}. Finally, the proposed approach is evaluated in Section~\ref{exp}.

\subsection{Dataset}
\label{dataset}
\begin{figure}
	\includegraphics[width=0.47\textwidth]{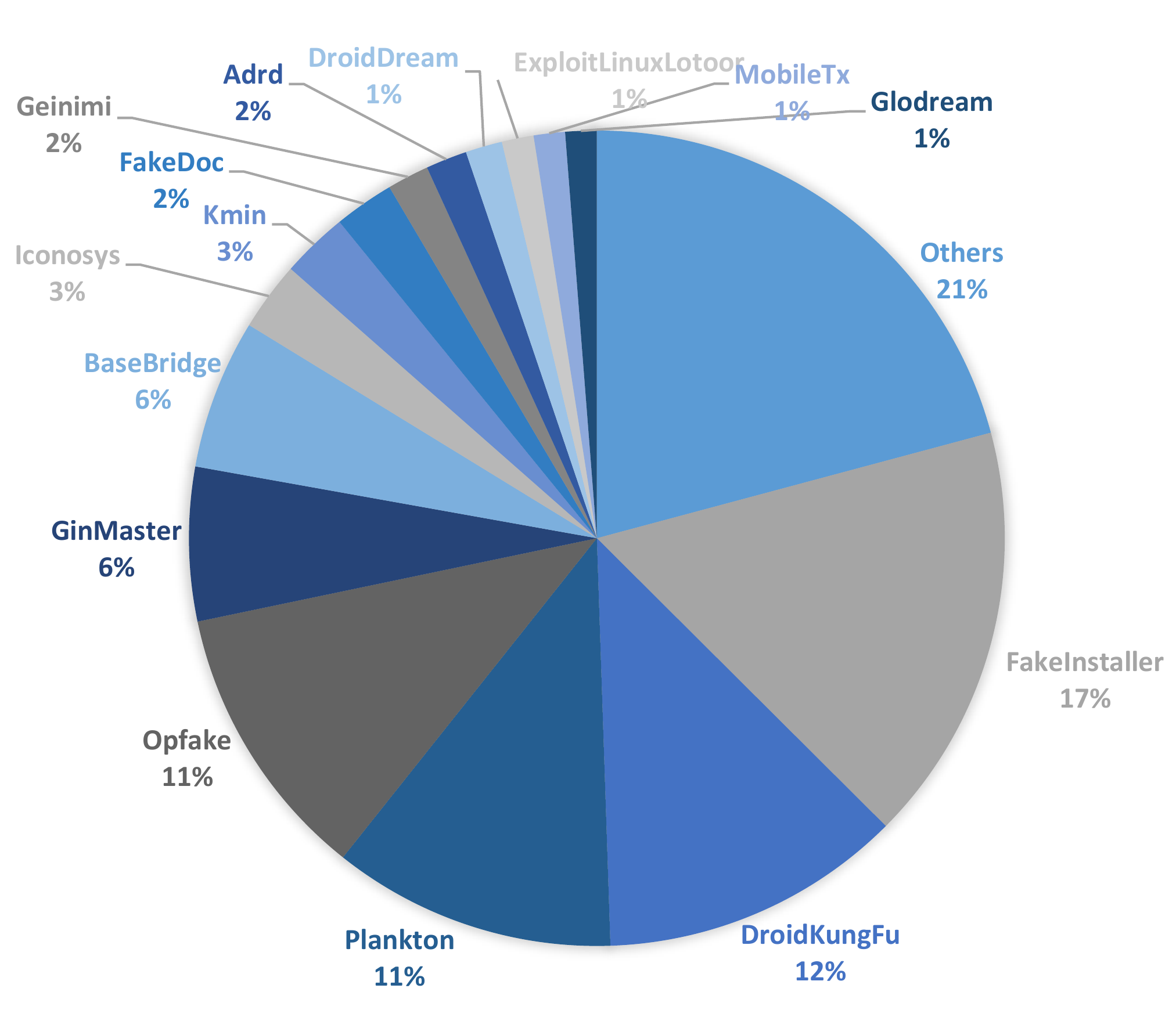}
	\caption{Distribution of samples in each malicious family in our initiated dataset.} 
	\label{plot:family}
\end{figure}

In order to evaluate the accuracy and performance of ServiceMonitor, we built a dataset of applications including benign and malicious samples.
Our dataset of malicious applications was composed of 5560 samples, from 179 different families, obtained from the Drebin dataset~\cite{drebin}, which  extends the Android Malware Genome Project dataset \cite{survey2}. The distribution of samples in each malicious family is illustrated in Fig. \ref{plot:family}. Note that the Drebin dataset was built for a static analysis approach. Therefore, a number of issues arise when employing dynamic analysis approaches on the samples provided in this dataset. First, a number of samples in Drebin dataset could not be installed on an Android device. More specifically, we found that 429 samples have broken APK files. Second, we were unable to execute 1097 samples for analysis, either due to the fact that the application was dependent on the presence of another application, or the application executed as a background service. Considering the noted issues with our initial dataset, we excluded such applications from the dataset, hence our malicious set contained a total of 4034 samples.

Furthermore, we crawled the official Android  store to collect 10370 samples from May 2016 to June 2016. In order to make sure that these samples were benign in nature, we submitted them to  VirusTotal\cite{virustotal}, which tests applications by fifty-four anti-malware engines. We found and eliminated 346 suspicious samples, therefore our benign dataset consisted of 10024 total number of samples.

\subsection{Feature Collection}
\label{fcollect}

In order to collect the described feature vectors from the described dataset, applications were installed on an unmodified Android version 4.4.4 (KitKat version), which was deployed upon a specific virtual machine called VirtualBox \cite{virtualbox} for executing and tracing them simultaneously. As mentioned earlier in this paper, the tracing of applications' behaviors with the aim of extracting the features are done through the  ServiceMonitor's kernel module.

In order to process the large number of applications in the dataset, we automated the process of installation, execution, interaction, and data collection. This was done by installing each application in a clean state of the operating system automatically by using $adb$ tool and simulating end users' activities and interactions. We leveraged  \emph{MonkeyRunner} tool \cite{monkey} in a script written in Python to simulate the interaction (e.g. screen clicks and touches) of end users with the system and application. By this script we were sending random events to the application with 2 milliseconds pause period between the successive events. Finally, after the execution of all events in the virtual machine, we stopped the execution of application and revert the machine to the clean state by replacing it with the clean snapshot of the virtual machine that was taken before installing the application. Taking a snapshot and reverting to it afterwards are functionalities of VirtualBox and we used them by implementing a bash script.

\subsection{Experimental Evaluation}
\label{exp}

In order to evaluate the effectiveness of ServiceMonitor, we employed the  Random Forests \cite{randomforest_imp} classification algorithm. For this experiment, we trained the classifier with the feature vectors obtained from Markov chain modeling procedure and also we used a $k$-fold cross-validation procedure with $k$=10 to overcome the over fitting problem. Furthermore, we should note that we extracted 51077 features for each application in the dataset. 
So, due to the machine learning problems with high dimensional data, it was difficult to have an efficient and accurate estimator in this case.
As a remedy, in this step we used a well-suited feature selection method called Principal Component Analysis (PCA)  \cite{pca}.
In this statistical procedure, PCA ranks features by considering their variance in the feature space. In other words, we apply PCA to identify uncorrelated and most important features to reduce the dimension of feature space and improve the efficiency and accuracy of the detection system as well.  
After application of PCA, we reduced the dimension to 200 components/features with the highest contribution to the classification decision (i.e. features with maximum variance). Finally, we used the reduced-dimension feature vectors for learning the classifier and building the detection model.

\begin{figure}
	\center
	\includegraphics[width=0.47\textwidth]{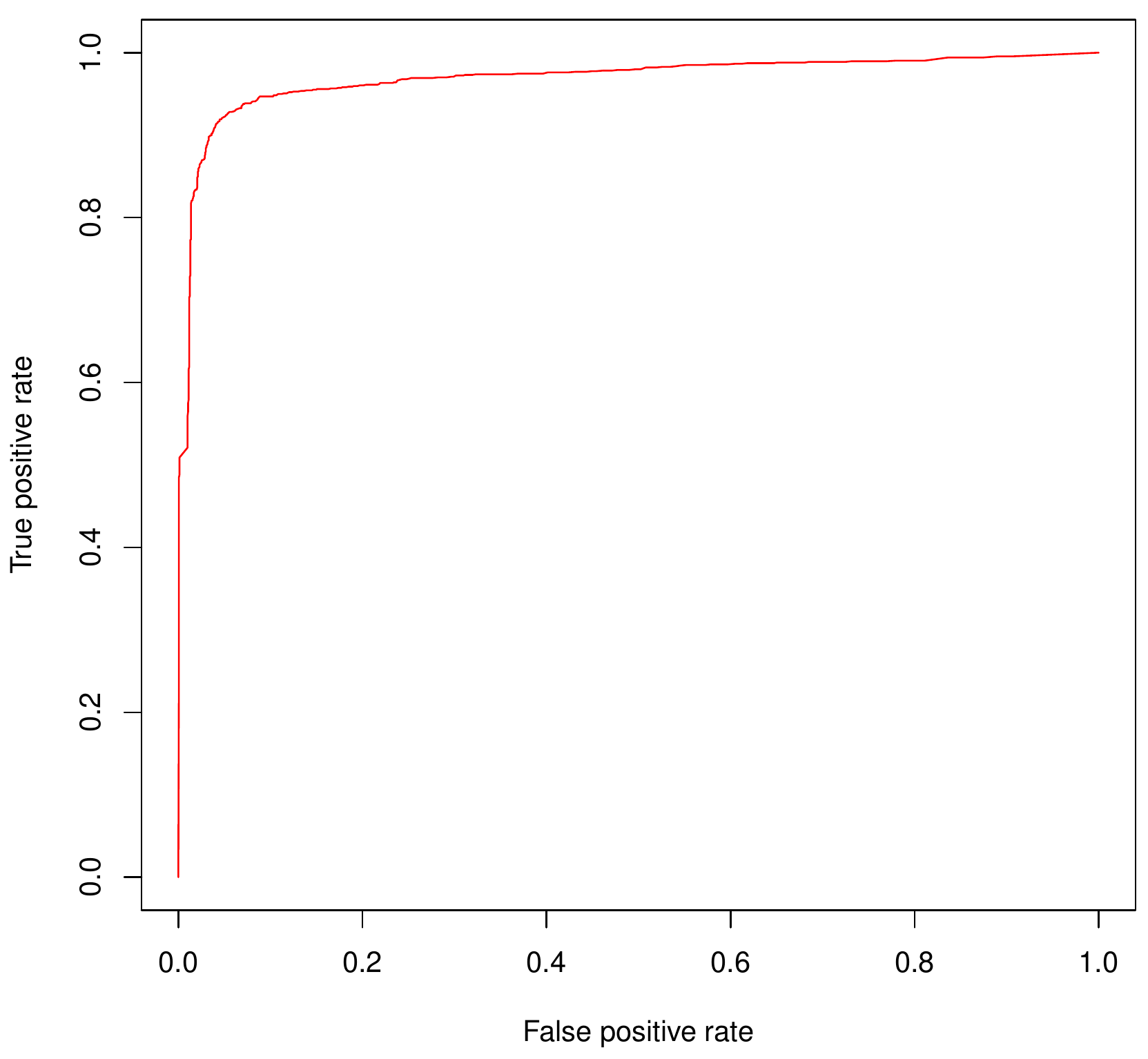}
	\caption{ROC curve obtained by evaluating the trained model, against the test dataset. The area under the ROC curve is 0.97.} 
	\label{plot:roc}
\end{figure}

The result of this experiment was encouraging. ServiceMonitor effectively classified the malicious and benign applications in the introduced dataset with accuracy rate of 96\%, false-negative rate of 2.1\%, and false-positive rate of 4.4\%. Fig. \ref{plot:roc} depicts the ROC curve of this experiment to illustrate the true-positive rate against different false-positive values. The calculated area under the ROC curve (i.e. AUC) is 0.97.

\subsubsection{Runtime Measurement}
\label{runtime}
With the aim of measuring the overhead of ServiceMonitor, we used the PassMark PerformanceTest v2.0 \cite{passmark} to benchmark the CPU and memory. In fact, PassMark conducts eight different tests to determine a device's PassMark rating for CPU and memory. Since the benchmark results of PassMark test are returned as indexes, the higher value in this test means better performance. As illustrated in Table \ref{tab:performance}, the overhead of ServiceMonitor on CPU and memory is acceptable. Actually, there is 0.8\% and 2\% performance impact on CPU and memory respectively. As shown, the highest overhead is on memory and this is mainly due to the IPC dissecting procedure that incurs overhead on Binder transactions.

\begin{table}[t]
	\centering
	\begin{threeparttable}
	\renewcommand{\arraystretch}{1.3}
	\caption{Run time overhead of ServiceMonitor, measured by passmark benchmark.}
	\label{tab:performance}
	\begin{tabular}{|c|c|c|c|}
		\hline
		\textbf{Test} & \textbf{Baseline} & \textbf{ServiceMonitor} &  \textbf{Overhead}  \\
		\hline
		\hline
		CPU&13520&13410&0.8\%\\
		\hline
		Memory&13860&13550&2\%\\
		\hline
	\end{tabular}
	\begin{tablenotes}
	
		\item Higher rating value means better performance.
	\end{tablenotes}
\end{threeparttable}
\end{table}

\subsubsection{Comparison with Related work}
Among the many proposals in the literature, we considered Patronus \cite{patronus} as the state of the art host-based approach, with which we could compare the proposed ServiceMonitor. However, as the Patronus implemented system was not available, we were not able to evaluate Patronus based on the dataset employed in our evaluations. Hence and in order to conduct a valid comparison, we employed a dataset with similar malicious families (i.e. BaseBridge, FakeAV, and MobileTx) as used in Patronus, to evaluate ServiceMonitor.  

With the similar dataset, ServiceMonitor obtains an accuracy rate of 97.5\% roughly 10\% higher than Patronus; Furthermore, ServiceMonitor has false-positive rate of 0\% against the Patronus's dataset, which is 1\% lower than Patronus's FPR. More importantly, Patronus is a policy-based solution, which is built on manually crafted malicious policies. Hence in contrast to ServiceMonitor, Patronus is not able to detect unknown malicious behaviors and zero-day malwares.  Also we should note that the overhead of ServiceMonitor is smaller than Patronus in the evaluated benchmark tests. Specifically, Patronus has 0.9\% and 8\% overhead on CPU and memory respectively in comparison with ServiceMonitor that has 0.8\% and 2\% overhead on CPU and memory respectively.

\section{Discussion}
\label{sec:Discussion}

In what follows, we discuss a number of observations made with respect to the different functionalities requested by the benign and malicious applications existing in the evaluation dataset. Furthermore, we discuss the limitations of the proposed service monitor technique for malware detection in Android.

\subsection{Observations}
Given the large dataset employed in the evaluation of the proposed technique, we were able to obtain further insight into how applications, either benign or malicious,  request access to different functionalities. 

\textit {1)	Telephony Manager:} As expected, requesting functions from telephony system services, like functions related to retrieve phone number and unique device ID (i.e. IMEI), occurs more commonly in malware applications. More specifically, more than 67\%  of the malware applications retrieve phone-related subscriber information, while there are only 9\% of the benign applications with similar functionalities.
Actually, telephony related methods like \emph{getDeviceId()}, \emph{getLine1Number()}, \emph{getSubscriberId()}, \emph{getIccSerialNumber()} are widely used in malicious applications in comparison with benign ones.

In addition, a great number of current malicious Android applications (like malware samples belong to OpFake and Gemini families) subscribe to premium-rate services and send SMS messages to them with profitability objectives. ‌Based on the results, while 17\% of the malicious applications have telephony activities which cause financial charges to the infected users, it was done by none of benign applications in our dataset. 

\textit {2)	Location Manager:}  Some malicious applications have access to GPS modules with the goal of collecting location data. For example, a malware family named AccuTrack is a family of applications that track down the GPS location of the device and turns it into a GPS tracker. In practice during the test, 12\% of the malware applications gained access to the location data, while only  7\% of the benign applications requested these services. 

\textit {3)	Package Manager:} In our experiments, it is common that malwares (like some samples in DroidDream malware family) invoke \emph{getInstalledPackages()} method to retrieve a list of all packages that are installed on the device to take an appropriate action. The result shows that about 16\% of the malicious applications requested such data during their test time while only 1\% of the benign applications requested it.

\textit {4)	Activity Manager:} \emph{getRunningAppProcesses()} method from the \emph{ActivityManager} class was invoked in 3.7\% of the malicious applications to retrieve the list of running application processes; however, only 1\% of the benign applications invoked this method. Applications can use this capability to check the presence of a running specific service (e.g., Anti-malware) in the device to take an action (like killing the anti-malware process). 
\emph{getMemoryInfo()} is another method used frequently by 5.7\% of the malicious applications to retrieve available memory space on the device, whereas only 0.7\% of the benign applications requested this method.

\textit {5) Service Manager:} 
As a whole, the results show that requesting system services in the malware applications has more frequency than the benign applications, such that on average each malware requested 38 system services during its test time whereas less than 12 service requests occurred in each benign application.  
Actually, we considered the occurrences of \emph{getService()} method, from the \emph{ServiceManager} class, in Android applications to determine the average of service requests in them.

\subsection{Limitations}
Even though ServiceMonitor is able to detect Android malwares accurately, we should note a number of issues which could limit its accuracy.
As noted in \cite{evading}, detecting the virtualization or emulation environments is one of the most popular methods employed by Android malware families (e.g. \textit{Android.HeHe} \cite{hehe} and \textit{OBAD} \cite{obad}) to evade analysis procedure and alter their behaviors accordingly. Therefore, due to the execution environment of ServiceMonitor's training phase, which is based on a specific virtual machine, some malwares could fingerprint the virtualized environment and avoid requests to the system services and hence be classified as benign in our experiments. Furthermore, some malicious samples were unable to show their malicious behavior, because for example the $C\&C$ servers of some malwares were not available during the analysis time or the malicious logic maybe hidden and only executed, or triggered, under specific circumstances.

As a learning-based method, ServiceMonitor might also be vulnerable to pollution and mimicry attacks \cite{mimicry}. In other words, malicious applications could randomly request system services and functionalities to change the original pattern of their requested functions from system services aim at confusing the detection system.

And finally, since loading the kernel module of ServiceMonitor requires root access, our detection system only runs on rooted devices. But the encouraging results of the ServiceMonitor for detecting malicious applications and its reasonable performance on mobile devices with limited resources, rendering it suitable for manufacturers to integrate ServiceMonitor directly on the operating systems of mobile devices.

\section{Conclusions}
\label{sec:Conclusion}
In this paper, we proposed a system service use analysis technique that systematically extracts fine-grained behaviors of applications based on their accesses to system resources. Furthermore, we designed and implemented a host based system, called ServiceMonitor, that dynamically tracks execution behaviors of applications based on the proposed system service use analysis technique and models these behaviors in the form of Markov chains to classify applications into benign or malicious. 

Our evaluation results show that ServiceMonitor is able to detect Android malwares accurately and efficiently on mobile devices. Employing the Random Forests
classifier against 4034 malwares and 10024 benign applications, ServiceMonitor were able to obtain the accuracy rate of 96\% in distinguishing malicious applications from the benign ones.

\ifCLASSOPTIONcompsoc

\ifCLASSOPTIONcaptionsoff
  \newpage
\fi



%
\bibliographystyle{IEEEtran}	
\bibliography{resources}

	
	

%

\end{document}